\documentclass[a4paper]{jpconf}
\usepackage{graphicx}

\begin{document}

\title{Closing in on the sources of Galactic and extragalactic cosmic rays\footnote{Written in celebration of the 75th birthdays of Frank Avignone and Ettore Fiorini and in memory of the 75th birthday of the late Peter Rosen.}}

\author{Francis Halzen}

\address{Department of Physics, University of Wisconsin,
Madison, WI 53706, USA}

\ead{Francis.Halzen@icecube.wisc.edu}

\begin{abstract}We speculate that a new generation of particle astrophysics instrumentation will reveal the enigmatic sources of cosmic rays prior to the one hundredth anniversary of their discovery by Hess in 1912. While only a ``smoking gun'' is missing for the case that the galactic component of the cosmic ray spectrum originates in supernova remnants, deciphering the origin of the extragalactic component is still at a level of reading tea leaves.
\end{abstract}


\section{The energetics of cosmic ray sources}

Cosmic accelerators produce particles with energies in excess of $10^8$\,TeV; we still do not
know where or how \cite{Sommers:2008ji}. The flux of cosmic rays observed at Earth is shown in Fig.\,1. The energy spectrum follows a broken power law. The two power laws are separated by a feature dubbed the ``knee'' at an energy\footnote{We will use energy units TeV, PeV and EeV, increasing by factors of one thousand from GeV-energy.} of approximately 3\,PeV.  There is evidence that cosmic rays up to this energy are Galactic in origin.  Any association with our galaxy disappears in the vicinity of a second feature in the spectrum referred to as the ``ankle"; see Fig.\,1. Above the ankle, the gyroradius of a proton in the Galactic magnetic field exceeds the size of the Galaxy and it is routinely assumed that we are witnessing the onset of an extragalactic component in the spectrum that extends to energies beyond 100\,EeV. Direct support for this assumption now comes from two experiments \cite{Abraham:2008ru} that have observed the telltale structure in the cosmic ray spectrum resulting from the absorption of the particle flux by the microwave background, the so-called Greissen-Zatsepin-Kuzmin cutoff\footnote{The possibility has not been eliminated that the ``cutoff" is simply the maximum energy reached by the accelerator(s).}. The origin of  the flux in the intermediate region covering PeV to EeV energies remains a complete mystery. Although the routine assumption is that it results from some high energy extension of the reach of the Galactic accelerators, no convincing mechanism for this has been identified.

\begin{figure}

\begin{center}
\includegraphics[width=6in]{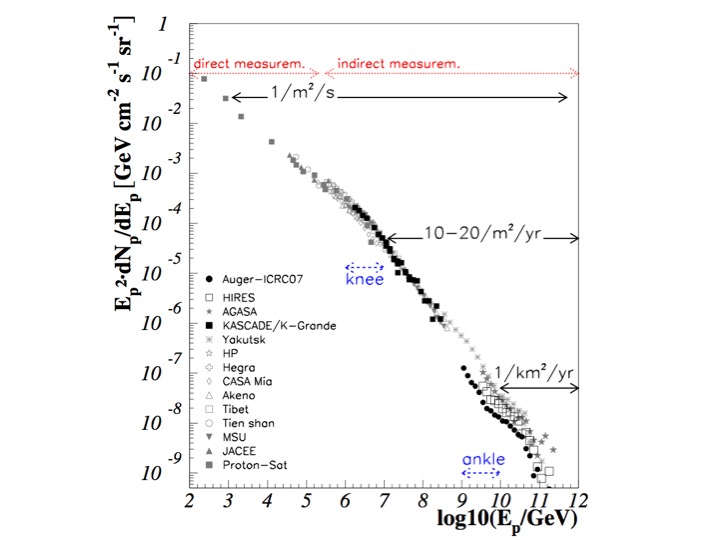}
\end{center}
\caption{At the energies of interest here, the cosmic ray spectrum follows a sequence of 3 power laws. The first two are separated by the ÒkneeÓ, the second and third by the ÒankleÓ. The cosmic rays beyond the ankle are a new population of particles produced in extragalactic sources.}
\end{figure}

It is assumed that cosmic rays originate in cosmic accelerators\footnote{Speculations on the origin of the highest energy cosmic rays fall into two categories,
top-down and bottom-up. In top-down models it is assumed that the cosmic rays
are the decay products of cosmological remnants or topological defects associated with, for instance, Grand Unified theories with unification energy $M_{\textrm{GUT}}\sim 10^{24}$\,eV.
These models predict, besides cosmic rays, large fluxes of gamma rays and neutrinos that have not been observed.}. Acceleration to TeV energy and above requires massive bulk flows
of relativistic charged particles. These are likely to originate from the exceptional
gravitational forces in the vicinity of black holes or neutron stars. Gravity powers large currents of charged particles that are the origin of high magnetic fields. These create the opportunity for particle acceleration by shocks. It is a fact that
black holes accelerate electrons to high energy; astronomers observe them indirectly
by their synchrotron radiation. Some must accelerate protons because we
detect them as cosmic rays.

The Swiss astronomer Fritz Zwicky suggested as early as 1933 that supernova remnants could be sources of the Galactic cosmic rays. It is assumed that the accelerators are powered by the conversion of $10^{50}$\,erg of energy into particle acceleration by diffusive shocks associated with young ($\sim\,1000$ year old) supernova remnants expanding into the interstellar medium \cite{ADV94}. Like a snowplough, the shock sweeps up the $\sim\,1\,\rm proton/cm^3$ density of hydrogen in the Galactic plane. The accumulation of dense filaments of particles in the outer reaches of the shock, clearly visible as sources of intense X-ray emission, are the sites of high magnetic fields. It is theorized that particles crossing these structures multiple times can be accelerated to high energies following an approximate power-law spectrum $dN/dE \sim E^{-2}$. The mechanism is familiar from solar flares where filaments of high magnetic fields accelerate nuclear particles to tens of GeV. The higher energies reached in cosmic ray accelerators are the consequence of particle flows of much larger intensity powered by the gravitational energy of collapsed objects such as neutron stars and black holes.

From a myriad of ideas, speculations on the sites for the acceleration of extragalactic cosmic rays have converged on the supermassive black holes at the centers of active galactic nuclei (AGN) or the primary engines of gamma ray bursts (GRB). It should surprise nobody, however, if the final answer turned out to be neither of these.

AGN are of special interest because some emit most of their luminosity at TeV energy and above. Their inferred isotropic luminosities can be as high as $10^{45}-10^{49}\rm{erg}\,\rm{s^{-1}}$. They produce a typical spectrum $dN/dE_\gamma \propto E^{-2}$ in the MeV-GeV range. The energetics requires mass accretion on a black hole that is up to one billion times more massive than our Sun. Some of the matter falling onto the black hole is deflected and accelerated in highly beamed jets aligned along its rotation axis. Both the inflow onto the central engine and the jet provide opportunities for the buildup of large magnetic fields and shock acceleration. For all these reasons AGN were pinpointed by Ginzberg and Syrovatskii  \cite{ginzberg} as far back as 1964 as candidate cosmic ray accelerators. A subset, called blazars, emit high-energy radiation in collimated jets pointing at the Earth and are the sources of photons with energies of tens of TeV. Their emission is highly variable over several time scales. TeV-energy bursts as short as minutes have been observed. Exceptionally, the nearest active galaxies, Centaurus A (Cen A) and M87 have been detected in TeV gamma rays even though their jets are not pointing at us.

Some argue instead that gamma ray bursts (GRB), outshining the entire Universe for the duration of the burst, are the best motivated sources of high energy cosmic rays \cite{waxmanbahcall}. The collapse of a massive star to a black hole has emerged as the origin of ``long" GRB with durations of tens of seconds. In the collapse a fireball is produced of electromagnetic plasma that expands with a highly relativistic velocity powered by radiation pressure. The fireball eventually runs into the stellar material that is still accreting onto the black hole. If it successfully punctures through this stellar envelope the fireball emerges to produce a GRB display. While the energy transferred to highly relativistic electrons is observed in the form of synchrotron radiation, it is a matter of speculation how much energy is transferred to protons. The assumption that GRB are the sources of the highest energy cosmic rays determines the energy of the baryons in the fireball. Accommodating the observed energy spectrum of extragalactic cosmic rays leads to the requirement of roughly equal efficiency for conversion of fireball energy into the kinetic energy of protons and electrons.

It is routinely emphasized that the flux of cosmic rays, especially at the highest energies, is very low. For example, at the onset of the extragalactic component near 10\,EeV the flux is only at the level of one particle per kilometer squared per year for a typical array with a steradian acceptance in angle. This can be translated into an energy flux
\begin{equation}
E \left\{ E\frac{dN}{dE} \right\} = \frac{10^{19}\,{\rm eV}} {\rm (10^{10}\,cm^2)(3\times 10^7\,s) \, sr} = 3\times 10^{-8}\rm\, GeV\ cm^{-2} \, s^{-1} \, sr^{-1} \,.
\label{crflux}
\end{equation}
The particles may be few, but each carries an enormous energy that can be expressed in macroscopic units, tens of Joules for the highest energies. We can derive the average energy density $\rho_{E}$ of cosmic rays in the Universe using the relation that the total flux${}={}$velocity${}\times{}$density, or
\begin{equation}
4\pi \int  dE \left\{ E{dN\over dE} \right\} =  c\rho_{E}\,.
\end{equation}
We obtain
\begin{equation}
\rho_{E} = {4\pi\over c} \int_{E_{\rm min}}^{E_{\rm max}} {3\times 10^{-8}\over E} dE \, {\rm {GeV\over cm^3}} \simeq 10^{-19} \, {\rm {TeV\over cm^3}} \,,
\end{equation}
taking the extreme energies of the accelerator(s) to be $E_{\rm max} / E_{\rm min} \simeq 10^3$. This is the corresponding value for cosmic rays of the energy density of, for instance, microwave photons of 410 photons of 2.7\,K per cubic centimeter.\footnote{We note that the energy density of photons changes by less than 5 orders of magnitude over the electromagnetic spectrum, from radio waves to GeV-photons, while the flux drops by almost 20 orders of magnitude.}

The energy content derived ``professionally" by integrating the observed extragalactic spectrum in Fig.\,1, including the GZK feature, is $\sim\,3 \times 10^{-19}\rm\,erg\ cm^{-3}$ \cite{TKG}. This is within a factor of our back-of-the-envelope estimate (1\,TeV = 1.6\,erg).

It has been realized for a long time that the corresponding quantity for the Galactic component of the spectrum may be a revealing number. If one repeats the integration that we just introduced for the Galactic flux in Fig.\,1, from GeV-energy, below which it is shielded from us by the sun, up to the knee near 3\,PeV, one obtains an energy density of the cosmic rays in our Galaxy of
\begin{displaymath}
\rho_{E} \sim 10^{-12}\,\textrm{erg}\,\textrm{cm}^{-3}.
\end{displaymath}
This happens to be very close to the energy density of light in our Galaxy and to the energy density $B^2/8\pi$ in its microgauss magnetic field. This density, as well as the one for extragalactic cosmic rays in the Universe, will represent informative benchmarks for speculating on the sources.

\section{Multi-wavelength Astronomy: Cosmic rays, gamma rays and neutrinos}

Although the origin of cosmic rays remains a matter of speculation almost one century after their discovery, particle astrophysicists have developed ambitious instrumentation that probes their sources with increased sensitivity detecting cosmic rays, gamma rays and neutrinos. In the 3-prong attack on the cosmic ray problem, TeV-astronomy is by far the most mature \cite{Krawczynski:2007dy}. A new generation of ground based air Cherenkov detectors has revealed plausible candidate cosmic ray sources that had been, surprisingly, not previously identified at other wavelengths. However, the basic hurdle to conclusively identify the observed TeV gamma rays as the decay products of pions produced by a cosmic ray accelerator has not been overcome. Synchrotron radiation by energetic electrons, routinely observed in non-thermal sources such as supernova remnants, AGN and GRB, cannot be excluded as their possible origin. Neutrinos from the charged pions accompanying pionic gamma rays can provide incontrovertible evidence for cosmic ray acceleration. The predicted fluxes are small and difficult to detect \cite{reviews}. Let's quantify the problem.

How many gamma rays and neutrinos are produced in association with the cosmic ray beam? For orientation,
consider a neutrino beam produced at an accelerator laboratory; see Fig.\,2. The accelerated protons interact with a target, referred to as a beam dump, producing pions that decay into gamma rays and neutrinos.\footnote{For earthbound accelerators the dump is designed to reabsorb all secondary electromagnetic and hadronic showers. Only neutrinos exit the dump. If Nature constructed such a ``hidden source'' in the heavens, conventional astronomy would not reveal it.} Generically, a cosmic ray source should also be a beam dump. Cosmic rays accelerated in regions of high magnetic fields near black holes inevitably interact with the radiation surrounding it, for instance, UV photons in active galaxies or the MeV-photons in GRB fireballs; see Fig.\,2. In these interactions they generate neutral and charged pions by the processes
\begin{eqnarray*}
p + \gamma \rightarrow \Delta^+ \rightarrow \pi^0 + p
\mbox{ \ and \ }
p + \gamma \rightarrow \Delta^+ \rightarrow \pi^+ + n.
\end{eqnarray*}
While the secondary protons may remain trapped in the high magnetic fields, neutrons and the decay products of neutral and charged pions escape. The energy escaping the source is therefore distributed among cosmic rays, gamma rays and neutrinos produced by the decay of neutrons, neutral pions and charged pions, respectively.

\begin{figure}[t]
\centering
\includegraphics[width=4in]{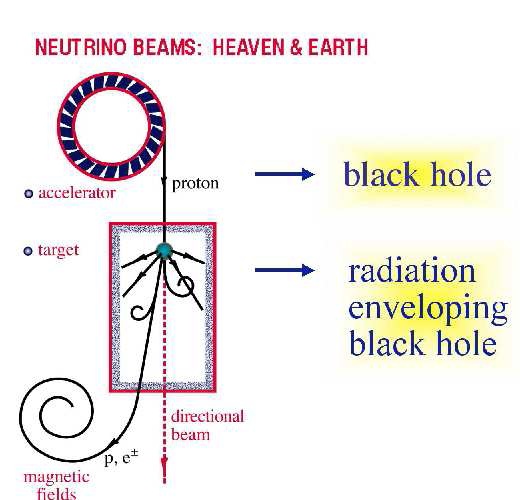}

\caption{Sketch of a cosmic ray accelerator producing photons and neutrinos. Cosmic rays accelerated in the vicinity of a black hole produce pions in interactions with the radiation and gases surrounding it. Neutral and charged pions decay into photons and neutrinos, respectively.}
\end{figure}

In reality the relation between the secondary fluxes is more complex. The cosmic rays may interact with gas besides radiation producing equal numbers of pions of all three charges in hadronic collisions $p+p \rightarrow n\,[\,\pi^{0}+\pi^{+} + \pi^{-}]+X$. Pionic gamma rays may cascade in the source steepening their spectrum and, as already pointed out, there may be an additional contribution to the gamma ray flux originating from purely electromagnetic processes.

The secondary flux of neutrinos and gamma rays in astrophysical sources can be related to the initial accelerated proton spectrum by specifying the final-state multiplicities of the pions and their average fractional energies $x_{i}$ relative to the proton. For instance, in hadronic collisions of cosmic rays interacting with gas in or near the source, equal numbers of each pion charge $\pi^{0}+\pi^{+} + \pi^{-}$ are produced. The neutral pions decay into two gamma-rays $\pi^{0} \rightarrow \gamma + \gamma$ and the charged pions into leptons and neutrinos $\pi^{\pm} \rightarrow e^{\pm} + \nu_{\mu} +\nu_{\mu} + \nu_{e}$. The effect of neutrino oscillations is to equalize the number of neutrinos of each flavor arriving at Earth, resulting into one neutrino of each flavor per charged pion. In summary, we obtain two-thirds of a charged pion per interacting proton and one neutrino (of each flavor) per charged pion; and one-third of a neutral pion per interacting proton and two photons per neutral pion. Therefore,
\begin{eqnarray}
\frac{dN_{\nu}}{dE}\left(E\right) = 1\times \frac{2}{3} \times \frac{1}{x_{\nu}}\times \frac{dN_{p}}{dE}\left(\frac{E}{x_{\nu}}\right) \\
\frac{dN_{\gamma}}{dE}\left(E\right) = 2\times \frac{1}{3} \times \frac{1}{x_{\gamma}} \times \frac{dN_{p}}{dE}\left(\frac{E}{x_{\gamma}}\right)
\end{eqnarray}

The fractional energies $x_{i}$ are determined by the average pion inelasticity measured to be approximately 0.2 at accelerators. In the approximation that the pion decay products carry equal energies, we obtain that $x_{\nu} \simeq 0.25\,\times x_{\pi}\simeq 0.05$ and $x_{\gamma} \simeq 0.5\,\times x_{\pi}\simeq 0.1$.

Notice that one can eliminate the initial cosmic ray flux to obtain a relation between secondary gammas and neutrinos. It just follows from the fact that there are 2(1) charged pions for every neutral pion in pp(p$\gamma$) interactions and, after taking into account oscillations, 2(8) photons for every $\nu_{\mu}$. The flux referred to sums over neutrinos and anti-neutrinos which the experiments cannot separate. For instance, for the case of a $E^{-2}$ spectrum elimination of the cosmic ray flux from above equations yields:
\begin{eqnarray}
\frac{dN_{\nu}}{dE}\left(E\right) = \frac{1}{4} \times \frac{x_{\gamma}}{x_{\nu}}\frac{dN_{\gamma}}{dE}\left(E\right)\simeq \frac{1}{2}\frac{dN_{\gamma}}{dE}\left(E\right).
\end{eqnarray}
In the photoproduction case the neutrino flux is further reduced by a factor 4.

\section{The Sources of Galactic Cosmic Rays}

We first recall that the energy density of the cosmic rays in our galaxy is $\rho_{E} \sim 10^{-12}$\,erg\,cm$^{-3}$. Galactic cosmic rays are not forever; they diffuse within the microgauss fields and remain trapped for an average containment time of  $3\times10^{6}$\,years. The power needed to maintain a steady energy density is therefore $10^{-26}\,\textrm{erg/cm}^{3}$s.  For a nominal volume of the Galactic disk of $10^{67}\,\textrm{cm}^{3}$ this requires an accelerator delivering $10^{41}$\,erg/s. This happens to be 10\% of the power produced by supernovae releasing $10 ^{51}\,$erg every 30 years.\footnote{$10 ^{51}\,$erg correspond to 1\% of the binding energy of a neutron star after 99\% is initially lost to neutrinos.} This coincidence is the basis for the idea that shocks produced by supernovae exploding into the interstellar medium are the accelerators of the Galactic cosmic rays.

Despite valiant attempts, it has been impossible to conclusively pinpoint supernova remnants as the sources of cosmic rays by identifying accompanying gamma rays of pion origin. An alternative signature of comic ray acceleration may prove to be more powerful. Additionally, pions are produced when the cosmic rays in the expanding remnant interact with the hydrogen in the Galactic plane. In the star forming regions where supernova are more likely to occur, the cosmic rays can interact with dense molecular clouds. We next estimate the TeV radiation from the Galactic plane associated with this indirect signature tracing cosmic ray acceleration.

A generic supernova remnant releasing an energy of $W\sim10^{50}\,$erg into the acceleration of cosmic rays will generate a source of TeV gamma rays by interacting with the hydrogen in the Galactic disk. The emissivity in pionic gamma rays $Q_{\gamma}$ is simply proportional to the density of cosmic rays $n_{cr}$ and the target density $n$ of hydrogen atoms. Here $n_{cr} \simeq 4\times10^{-14}\, {\rm cm}^{-3}$ by integrating the spectrum in Fig.\,1 for energies in excess of 1 TeV. For a $E^{-2}$ spectrum,
\begin{eqnarray}
Q_\gamma &\simeq& c \left<{E_\pi \over E_p}\right> {\lambda_{pp}}^{-1}\, n_{cr}\ ({>}1\,{\rm TeV}),\\
&\simeq& 2 c x_{\gamma}  \sigma_{pp}\, [n\, n_{cr}].\\
\noalign{\hbox{or}}
Q_\gamma (> 1\,{\rm TeV}) &\simeq& 10^{-29} \,{\rm photons\over \rm cm^3\,s}\, \left({n \over \rm 1\,cm^{-3}}\right).
\end{eqnarray}
The proportionality factor is determined by particle physics; $x_{\gamma}$ is the average energy of secondary photons relative to the cosmic ray protons and $\lambda_{pp}= (n\sigma_{pp})^{-1}$ is the proton interaction length ($\sigma_{pp} \simeq 40$\,mb) in a density $n$ of hydrogen atoms. The corresponding luminosity is
\begin{equation}
L_{\gamma} ({>} 1\,{\rm TeV})  \simeq Q_{\gamma}\, {W \over \rho_E} \simeq 10^{33}\, \rm photons\> s^{-1},
\end{equation}
where $W/\rho_E$ is the volume occupied by the supernova remnant. We here made the approximation that the volume of the young remnant is approximately given by $W/\rho_E$ or, that the density of particles in the remnant is not very different from the ambient energy density $\rho_E \sim 10^{-12}$\,erg\,cm$^{-3}$ of Galactic cosmic rays.

We thus predict a rate of TeV photons from a supernova at a nominal distance $d$ of order 1\,kpc~of
\begin{equation}
{E {dN_{\rm events}\over dE}}\,({ >}1 TeV) = {L_\gamma \over 4\pi d^2} \simeq 10^{-12}-10^{-11}  \left({\rm photons\over \rm cm^2\,s}\right) \left({W\over \rm 10^{50}\,erg}\right) \left({n\over \rm 1\,cm^{-3}}\right) \left({d\over \rm 1\,kpc}\right)^{-2}.
\end{equation}
Such sources must emerge in an all-sky TeV gamma ray survey performed with an instrument with the sensitivity of the Milagro experiment\, \cite{Abdo:2006fq}.

Above estimate ignores the fact that supernovae are not uniformly  distributed but are associated with regions of star formation such as the Cygnus region and the center of the galaxy. Of particular interest is the relatively close association of O, B stars in the Cygnus region at only $\sim 1$kpc. While the density of matter in the galactic plane is one proton per cubic centimeter, dense molecular clouds are common in star forming regions and can reach densities that are larger by two or three orders of magnitude. These are therefore efficient at converting cosmic rays into pions that decay into gamma rays and neutrinos.  They are an indirect signature for cosmic ray acceleration in nearby remnants and, unlike what is the case for the remnants themselves, there is no electromagnetic contribution to the TeV radiation that must be differentiated from the pionic gamma rays. The position of the knee in the spectrum indicates that some sources accelerate cosmic rays to energies of several PeV. These ``PeVatrons" therefore produce pionic gamma rays whose spectrum extends to several hundred TeV without cutoff in interactions with the interstellar medium, in particular with the dense clouds \cite{gabici}.

To date, the Milagro collaboration has identified 6 such PeVatron candidates. Three of these sources, MGROJ1908+06,
MGROJ2019+37, and MGROJ2031+41, have post-trial significances of $> 4.5\,\sigma$ (the only other Milagro source
of such statistical significance being the Crab nebula). Two of the remaining three hotspots, MGROJ2043+36 and MGROJ2032+37, are located within the Cygnus region of the Galaxy. The remaining hotspot, MGROJ1852+01, has a pre-trial significance of $5.1\,\sigma$; it is the strongest source but has the lowest significance, presumably because of large local background. One has to realize that, after subtracting the sources considered here, an excess of TeV gamma rays persists in the Milagro's skymap from the general direction of the Cygnus region \cite{Abdo:2008if}.

Of particular interest is MGROJ1908+06 which has been confirmed by the H.E.S.S. telescope. Their observations reveal a spectrum consistent with a $E^{-2}$ dependence from 500\,GeV to 40\,TeV without evidence for a cutoff\, \cite{hesshotspot}. In a follow-up analysis the MILAGRO collaboration\, \cite{UHEmilagro} showed that its own data are consistent with an extension of the H.E.S.S. spectrum to at least 90\,TeV; see Fig.~3. This is suggestive of pionic gamma rays from a PeVatron whose cosmic ray beam extends to the knee in the cosmic ray spectrum at PeV energies. Another source, MGROJ2031+4, has been observed \cite{magic2032} by the MAGIC telescope with a spectrum that is also consistent with $E^{-2}$. The lower flux measured by MAGIC can be attributed to the problem of differentiating the source from the background in a high density environment like the Cygnus region.

\begin{figure}[h]
\centering
\includegraphics[width=4.5in]{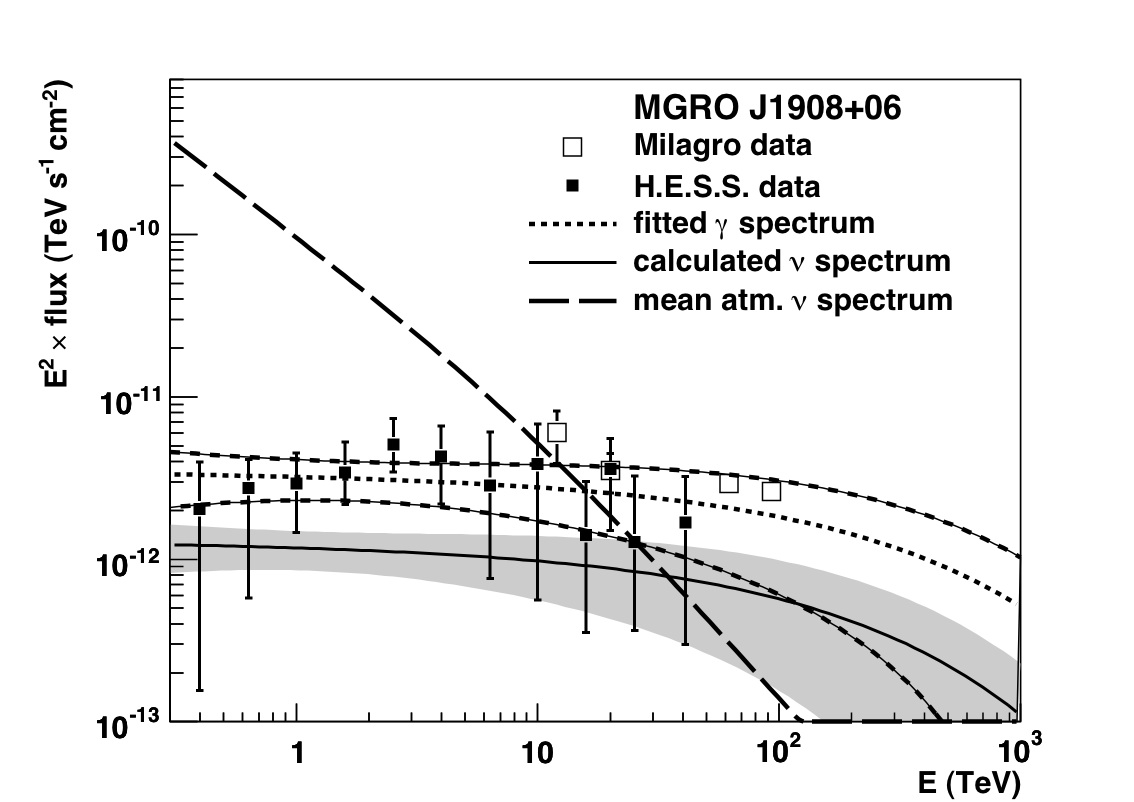}
\caption{The gamma ray and neutrino fluxes from MGROJ1908+06. The shading surrounding the neutrino flux represents the range in spectra consistent with the statistical and systematic uncertainties on the input gamma ray flux. Also shown is the flux of atmospheric neutrinos at the same zenith angle as the source (dashed line), taking into account the source size and angular resolution.}
\end{figure}

In the end, despite the suggestive evidence, conclusively tracing the observed gamma rays to pions produced by cosmic-ray accelerators has so far been elusive. It is one of the main missions of neutrino telescopes to produce the smoking gun for cosmic-ray production by detecting neutrinos associated with the charged pions. Neutrino telescopes detect the Cherenkov radiation from secondary particles produced by the interactions of high energy neutrinos in highly transparent and well shielded deep water or ice. They take advantage of the large cross section of high-energy neutrinos and the long range of the muons produced. The instrumentation consists of a lattice of photomultipliers deployed in a shielded and optically clear medium that is transformed into a Cherenkov detector. The IceCube telescope\, \cite{ice3} is under construction and is taking data with a partial array of 2400 ten inch photomultipliers positioned between 1500 and 2500 meter and deployed as beads on 40 strings below the geographic South Pole. With the completion of the detector by 2010-2011 the volume presently instrumented will be doubled from 0.5 to $1\,{\rm km^{3}}$.

As previously emphasized, particle physics is sufficient to compute the neutrino fluxes associated with pionic gamma rays. The neutrino flux associated with MGRO J1908+06 is shown in Fig.\,3 along with the TeV gamma ray observations from which it is derived.  The sensitivity of IceCube to the Milagro sources has been evaluated assuming that the 6 sources represent the imprint of the Galactic cosmic-ray accelerators on the TeV sky \cite{Halzen:2008zj}. While the number of events with energies of tens of TeV is relatively low, it is clear that this is the energy region where the atmospheric neutrino background, also shown in Fig.\,3, is suppressed and an excess from these sources can be statistically established. While observing individual sources may in some cases be challenging, establishing a correlation between the Milagro and IceCube sky maps should be conclusive after several years; see Fig.\,4. A ``stacked" source search that will look for correlations between all six Milagro sources and the IceCube sky map shown has a Poisson probability of 3--5\,$\sigma$ after 5 years; for a sample calculation see Fig.\,5. The range reflects the imprecise knowledge of the gamma ray fluxes. The use of optimised methods using unbinned searches beyond the simple binned method considered here will further increase IceCube's sensitivity.

\begin{figure}[h]
\centering
\includegraphics[width=4.5in]{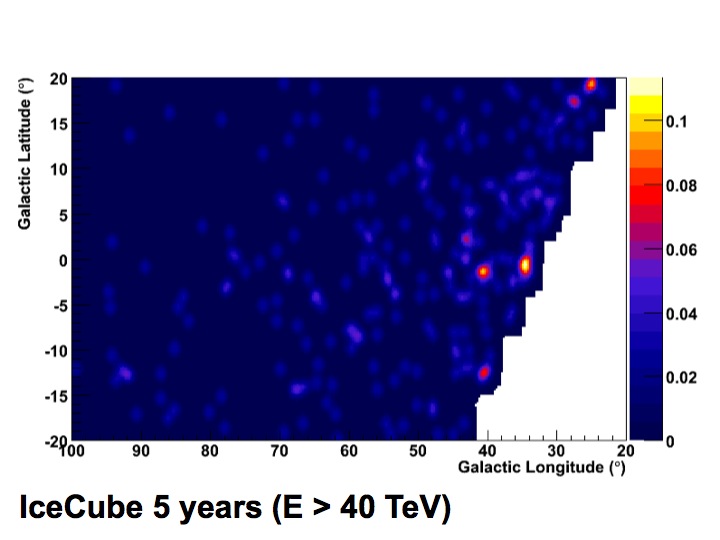}
\caption{Simulated sky map of IceCube in Galactic coordinates after 5 years of operation of the completed detector.  Two of the Milagro sources are visible ``by eye" with 4 events for MGROJ1852+01 and 3 events for MGROJ1908+06 with energy in excess of 40\,TeV. These as well as the background events have been randomly distributed according to the resolution of the detector and the size of the sources.}
\end{figure}

It is important to emphasize that the photon flux from the Milagro sources is consistent with the flux expected from a typical cosmic-ray-generating supernova remnant interacting with the interstellar medium. In other words, the TeV flux is consistent with the energetics that is required to power the cosmic ray flux in the Galaxy. Alternative sources such as microquasars have been theorized to contribute to the Galactic cosmic rays. If this were indeed the case, cosmic ray energetics would require that they leave their imprint on the Milagro skymap, but none have been observed so far.

\begin{figure}
\centering
\includegraphics[width=4.5in]{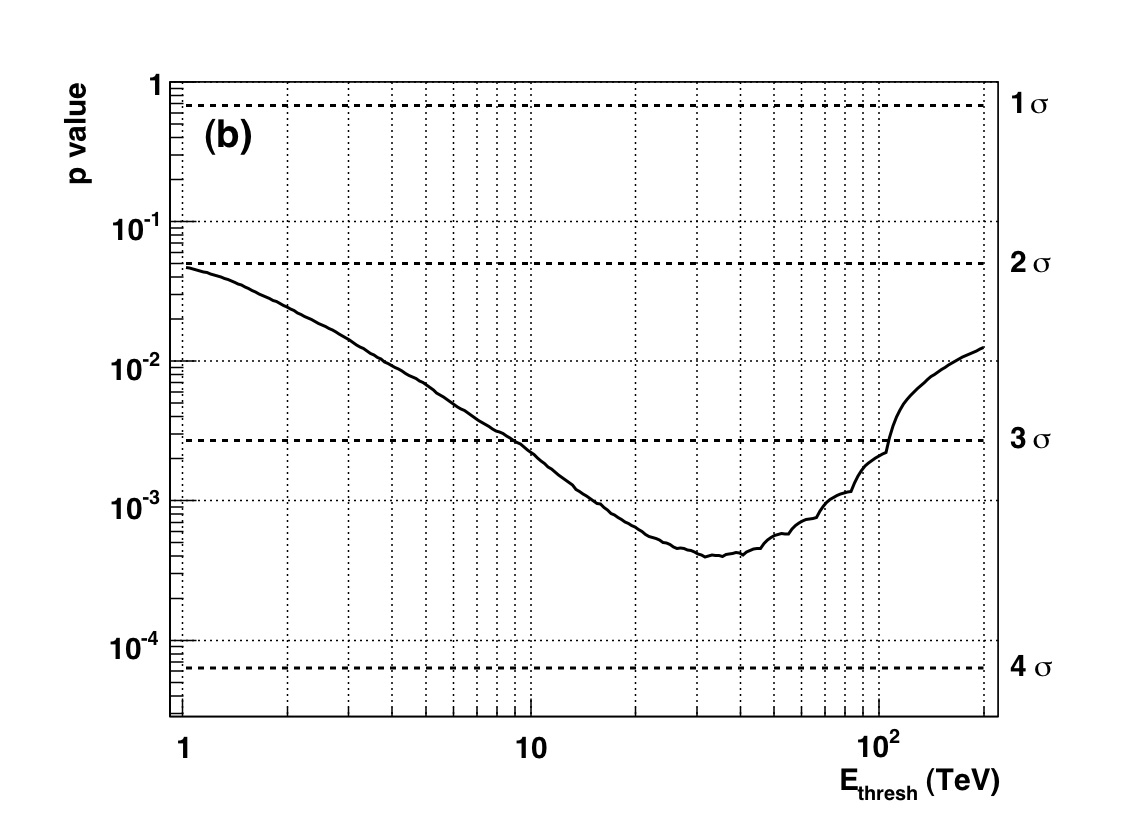}

\caption{Poisson probability of the excess from the 6 Milagro hotspots after 5 years as a function of the energy of the neutrino-induced muons and a gamma-ray cut-off of the source at 300TeV. Clearly most sensitivity is associated with muons with energies in excess of  40\,TeV.}
\end{figure}

\section{The Sources of the Extragalactic Cosmic Rays}

We first return to the observable that seemed so revealing in the case of the Galactic component, the energy density which is $\sim 3 \times 10^{-19}\rm\,erg\ cm^{-3}$ for extragalactic cosmic rays. The power required for a population of sources to generate this energy density over the Hubble time of $10^{10}$\,years is $\sim 3 \times 10^{37}\rm\,erg\ s^{-1}$ per (Mpc)$^3$ or, as often quoted in the literature, $\sim 5\times10^{44}\rm\,TeV$ per (Mpc)$^3$ per year.

As previously discussed, a GRB fireball converts a fraction of a solar mass into the acceleration of electrons, seen as synchrotron photons.  The energy in extragalactic cosmic rays can be accommodated with the reasonable assumption that shocks in the expanding GRB fireball convert roughly equal energies into the acceleration of electrons and cosmic rays. It so happens that $\sim 2 \times 10^{52}$\,erg per cosmological gamma ray burst will yield the observed energy density in cosmic rays after $10^{10}$ years given that their rate is of order 300 per $\textrm{Gpc}^{3}$ per year. Hundreds of bursts per year over Hubble time produce the observed cosmic ray density, just like 3 supernova per century accommodate the steady flux in the Galaxy. Problem solved? Not really, it turns out that the same result can be achieved with active galaxies.

The energy density of $3 \times 10^{-19}\rm\,erg\ cm^{-3}$ works out to not only \cite{TKG}
\begin{itemize}
\item $\sim 2 \times 10^{52}$\,erg per cosmological gamma ray burst, but also to
\item $\sim 3 \times 10^{42}\rm\,erg\ s^{-1}$ per cluster of galaxies,
\item $\sim 2 \times 10^{44}\rm\,erg\ s^{-1}$ per active galaxy.
\end{itemize}
The coincidence between above numbers and the observed output in electromagnetic energy of these sources explains why they have emerged as the leading candidates for the cosmic ray accelerators. Whether GRB or AGN, the observational evidence that the sources radiate similar energies in photons and cosmic rays  is consistent with the beam dump scenario previously discussed. In the interaction of cosmic rays with radiation and gases near the black hole, roughly equal energy goes into the secondary neutrons, neutral and charged pions whose energy ends up in cosmic rays, gamma rays and neutrinos, respectively.

Can IceCube reveal the extragalactic cosmic ray sources? Naively, the neutrino flux should be the same as the observed flux of cosmic rays of Eq.\,1. Not so naively, it is about 5 times smaller\footnote{For experts, it is the Waxman-Bahcall ``bound" adjusted downward because only 20\% of the proton energy is transferred to pions \cite{WB}.}; we will refer to the energetics estimate for the neutrino flux accompanying the extragalactic cosmic ray accelerators as the band shown in Fig.\,6, or
\begin{equation}
E_{\nu}^{2} dN / dE_{\nu}= 1- 5 \times 10^{-8}\rm\,GeV \,cm^{-2}\, s^{-1}\, sr^{-1}
\label{extragalactic}
\end{equation}
As illustrated in the figure, after 7 years of operation AMANDA's sensitivity is approaching the interesting range, but it takes IceCube to explore it.

\begin{figure}[h]
\centering
\includegraphics[width=4.5in]{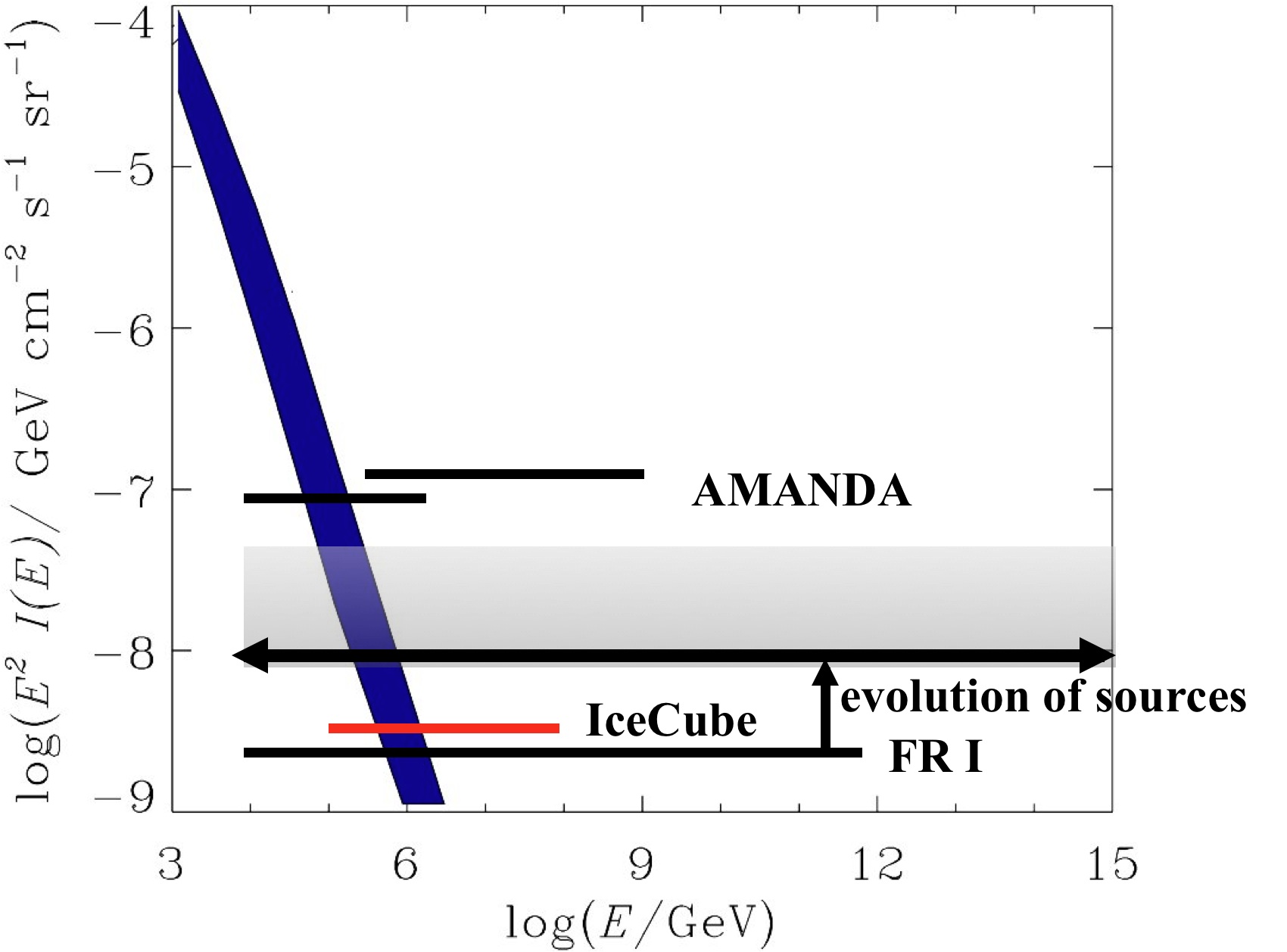}
\caption{Our energetics estimate of the flux of neutrinos associated with the sources of the highest energy cosmic rays (the shaded range) is compared to the limits established by the AMANDA experiment and the sensitivity of IceCube \cite{merida}. Also shown is the flux derived from the assumption that AGN are the sources which we model using spectral energy information on the nearby active galacties Cen A and M87; see Fig.\,7. Integration of AGN to larger redshifts can reconcile the two estimates which differ by a factor 3. Also shown is the background flux of atmospheric neutrinos.}
\end{figure}

\begin{figure}
\centering
\includegraphics[width=5.5in]{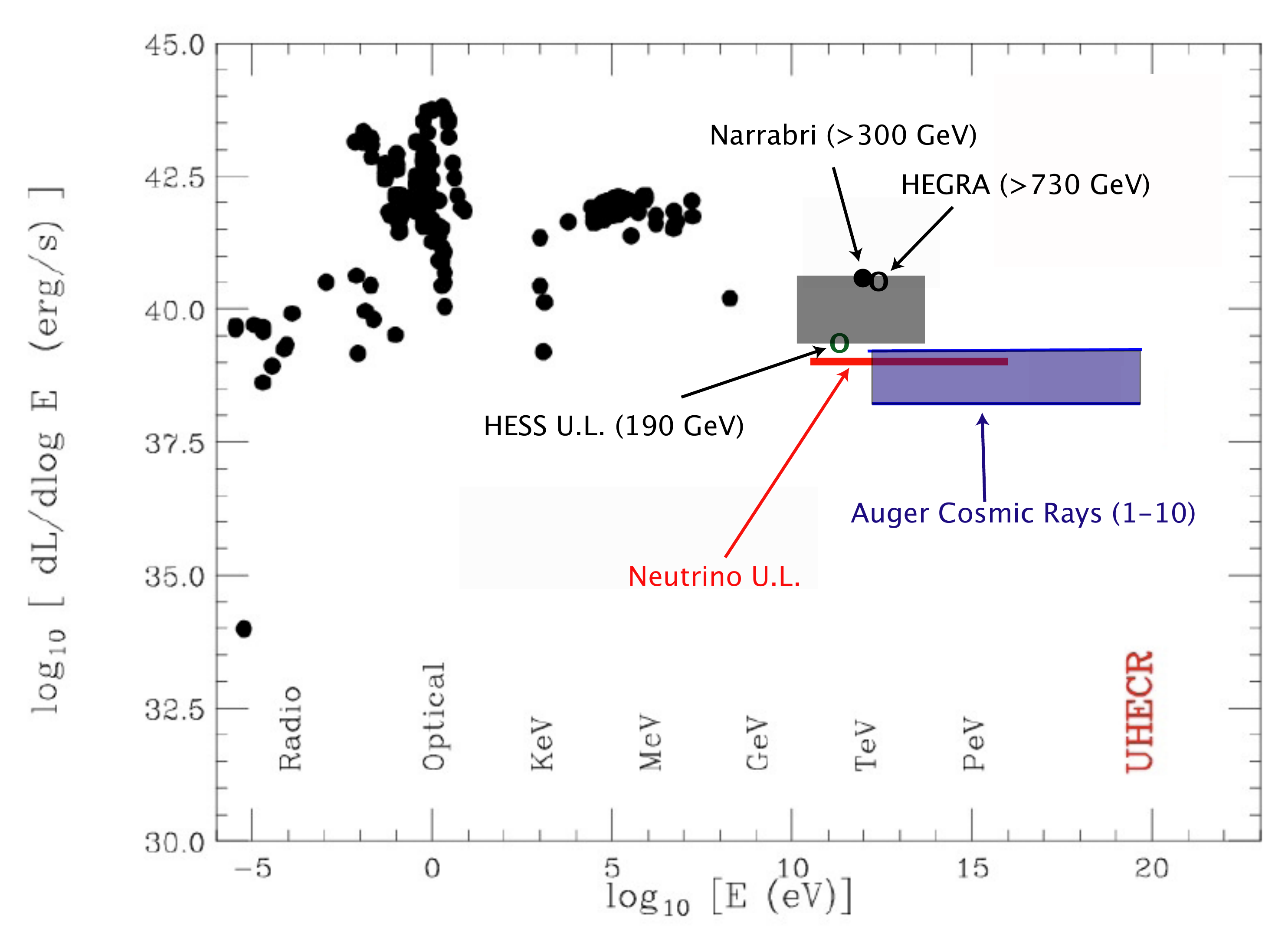}
\caption{Spectral energy distribution of Cen A (black dots). Having in mind that the source is variable, we show our estimates for the flux of TeV gamma rays (gray shading) and cosmic rays assuming that between 1 and 10 events observed by Auger originated at Cen A (blue shading). We note that the cosmic ray and TeV gamma ray fluxes estimated in this paper are at the level of the electromagnetic component shown from radio waves to GeV photons. Our estimate for the neutrino flux (labeled upper limit) is shown as the red line. }
\end{figure}

Recently the Pierre Auger observatory has presented evidence that the arrival directions of cosmic rays with energies in excess of $6\times10^{7}$\,TeV may be correlated with nearby active galactic nuclei\, \cite{auger}. The skymap suggests that the events cluster in the direction of the supergalactic plane\footnote{One should of course realize that any sources limited to a GZK absorption length of less than 100\,Mpc are likely to be clustered in the supergalactic plane.}, possibly in the direction of the FRI radio galaxy Centaurus A, by far the closest AGN at only 3.4\,Mpc. If AGN are indeed the sources, Cen A's proximity singles it out as a primary accelerator. We will review the information on its spectral energy distibution and argue that the observations are consistent, within large ambiguities, with the neutrino flux estimated above. In fact, Cen A and M87, the next nearest FRI whose jet is not aligned to our line of sight, have previously been singled out as potential cosmic ray accelerators on the basis of gamma-ray data\, \cite{anchordoquicena}.

The problem of interpreting TeV observations is however challenging because the high energy emission of AGN is highly variable and it is difficult to compare multi-wavelength data taken at different times. Extragalactic TeV sources observed so far are BLLacs, a subclass of Fanaroff-Riley I (FRI) active galaxies whose jets are oriented along our line of sight. The data have yielded no hints of cosmic ray acceleration so far; the observed spectra can be readily accommodated by synchrotron radiation by electrons, followed by inverse Compton scattering of the photons to TeV energy. Exceptionally, the nearby (16 Mpc) FRI M87 was observed in the late 1990's by the HEGRA stereoscopic system of five imaging atmospheric Cherenkov telescopes despite the fact that its jet is angled to more than $30^{\circ}$ from our line of sight\, \cite{Horiuchi}.

Where Cen A is concerned, there exists archival data of TeV emission collected in the early 1970's with the Narrabri optical intensity interferometer of the University of Sydney\, \cite{cenasydney}.
At the time the Sydney interferometer unsuccessfully searched the sky for gamma ray sources detecting the highest fluctuation in the direction of Cen A. As a followup they exclusively observed the source for a 3 year period. A variable flux was observed in two periods of heightened activity lasting roughly one year, pointing at a region of coherent emission of size of order 0.3 pc. This is consistent with the idea that the high energy emission is from an isotropic region near the base of the jet or the central black hole of mass $2\times 10^{8}$ solar masses, about two orders of magnitude more massive than the one at the center of our galaxy\, \cite{anchordoquicena}. The fact that the interferometer beam did not include the radio lobes at the end of the jets further supports the idea of a central engine at the base of the jets. An average flux of
\begin{equation}
dN_{\gamma}/d(\ln E) (E_{\gamma}  > 0.3\,{\rm TeV}) = 4.4 \pm 1.0 \times{\rm 10^{-11}\,cm^{-2}\,s^{-1}}
\end{equation}
was reported for the period of 3 years over which the source was variable.

The integrated flux of Cen A is close to the one observed by HEGRA\, \cite{hegra} from M87
\begin{equation}
dN_{\gamma} (E_{\gamma}/d(\ln E) > 0.73\,{\rm TeV}) = 0.96 \pm 0.23 \times{\rm 10^{-12}\,cm^{-2}\,s^{-1},}
\end{equation}
after scaling the flux of M87 at 16\,Mpc to the distance to Cen A and adjusting for the different thresholds of the experiments. The observations are therefore consistent with identical source luminosities of roughly $7\times10^{40}\, {\rm erg\,s^{-1}}$, assuming an $E^{-2}$ gamma-ray spectrum. This suggests that they may be generic FRI, a fact we will exploit to construct the diffuse neutrino flux from all FRI. Recently, MAGIC reported short day-long bursts of M87 \cite{magic} pointing at acceleration of particles even closer to the black hole. This, as well as the absence of associated X-ray activity expected in the case of electromagnetic processes, further supports the possibility of cosmic ray acceleration.

For completeness, the H.E.S.S. experiment obtained a limit on Cen A\, \cite{cenahess} of $10^{-12}$ in the same units, a flux smaller than the one just argued for. Given the burst nature of the data the disagreement may be acceptable. We have compiled the TeV gamma ray data in Fig.\,7  along with other observations on the multi-wavelength emission spectrum of Cen A recently compiled by Lipari \cite{lipari}.

The same conversion of TeV gamma rays to neutrinos, exploited for Galactic sources, yields a neutrino flux of
\begin{equation}
\frac{dN_{\nu}}{dE} \leq 5\times10^{-13} \left( \frac{E}{\rm TeV} \right)^{-2} {\rm TeV^{-1}\,cm^{-2}\,s^{-1},}
\end{equation}
conservatively normalized not to exceed the contemporary limit of the H.E.S.S. experiment.

Finally, having previously argued for a relation between the energetics of TeV gamma rays, cosmic rays and neutrinos, we ask the question what Auger data may reveal about the flux of the source. The answer depends on the number of events in the sky map, $N_{\rm events}$, that actually correlate with Cen A, a number that depends on the angular broadening of the source by the deflection of the particles by magnetic fields. $N_{\rm events}$ is therefore a matter of speculation. We estimate the flux assuming a power-law spectrum of the form $dN/dE =
N_0(E/E_0)^\alpha$, where the normalization $N_0$ is fixed by the number of events observed $N_{\rm events}$. For $\alpha = -2.0$, we have $N_{\rm events} = \mbox{field of view} \times \mbox{time} \times \mbox{efficiency} \times N_0/E_{\rm thresh}$. This gives us $dN_{\rm cr}/dE = N_{\rm events} \times 10^{-13}(E/\rm TeV )^{-2}\, TeV^{-1} \, cm^{-2} \, s^{-1}$. The number of events with energy in excess of $6\times10^{7}$\,TeV may be as many as 10, thus obtaining the range of cosmic ray fluxes shown in Fig.\,7. It is close to, but below the flux of TeV gamma rays and suggest an even smaller neutrino flux than the one derived from TeV gamma ray information. The variability of the source and the possibility of a more complicated shape of the spectrum clearly prevent us from reaching quantitative results on the basis of the present data.
 
The neutrino flux from a single source such as Cen A is clearly small: repeating the calculation for power-law spectra between 2.0 and 3.0, we obtain, in a generic neutrino detector of effective muon area $1\,{\rm km^{2}}$, between 0.8 and 0.02 events/year only. Having estimated the neutrino flux from a point cosmic ray source, we calculate next the total diffuse flux from all such sources within our horizon. Given an FRI density of $n \simeq 8\times 10^{4} \,{\rm Gpc^{-3}}$ within a horizon of $R\sim3 \,{\rm Gpc}$\, \cite{fridensity}, the total diffuse flux from all $4\pi \,{\rm sr}$ of the sky is simply the sum of the luminosities of the sources weighted by their distance:
\begin{equation}
\frac{dN_\nu}{dE_{\rm diff}} = \sum \frac{L_{\nu}}{4\pi d^{2}} =L_{\nu} \, n\,R = 4\pi d^{2} n R\frac{dN_\nu}{dE},
\end{equation}
where $\frac{dN_\nu}{dE}$ is given by Eq.\,1. We performed the sum by assuming that the galaxies are uniformly distributed. This evaluates to:
\begin{equation}
\frac{dN_\nu}{dE_{\rm diff}} = 2\times 10^{-9}\,\left(\frac{E}{\rm GeV}\right)^{-2}\,{\rm GeV^{-1}\,cm^{-2}\,s^{-1}\,sr^{-1}},
\end{equation}
approximately a factor 3 below the flux estimated on the basis of source energetics. Varying the spectral indices as before, we obtain an event rate per ${\rm km^{2}}$ year from the northern sky of between 19 and 0.5 neutrinos per year. Considering sources out to 3\,Gpc, or a redshift of order 0.5 only, is probably conservative. Extending the sources beyond $z\sim1$ increases the flux by a factor 3 or so as was discussed in connection with the results in Fig.\,6. We can thus bridge the gap between this and the previous estimate of the cosmic neutrino flux based on the energetics of the sources.

The predicted flux of Eq.\,16 should be within reach of IceCube and a future Mediterranean kilometer-scale neutrino telescope. The flux is close to IceCube's 90\% confidence level sensitivity in a single year. It follows that detection at the $5\sigma$ level should occur within $\sim 5$ years.\footnote{We here used the atmospheric background flux from charm particles predicted by\, \cite{prompt} rather than the larger flux conservatively implemented in the sensitivity calculation presented in the IceCube publication. The enhancement in the charm flux relative to the one used here is a direct consequence of the assumption that charm fragmentation functions scale and therefore clearly unphysical.}

It is not challenging to produce models that generate a neutrino flux at the level proposed here. For instance \cite{anchordoquicena}, the central engine in FRI such as Cen A and M87 may feed a beam dump consisting of the gas surrounding the supermassive black hole\footnote{Given the paucity of data, models can be invoked where, alternatively, the gamma rays are produced in photohadronic ($p\gamma$) rather than hadronic ($pp$) interactions\, \cite{hannestad}and in the jet rather than the black hole.}.

In summary, while the road to the identification of the sources of the Galactic cosmic ray has been mapped, the origin of the extragalactic component remains as enigmatic as ever.

\ack
I would like to thank my collaborators Alexander Kappes and Aongus \'O\,Murchadha as well as John Beacom, Julia Becker, Peter Biermann, Steen Hannestad and Stefan Westerhoff for valuable discussions. This research was supported in part by the National Science Foundation under Grant No.~OPP-0236449, in part by the U.S.~Department of Energy under Grant No.~DE-FG02-95ER40896, and in part by the University of Wisconsin Research Committee with funds granted by the Wisconsin Alumni Research Foundation.


\end{document}